\documentclass[8pt,a4paper,twocolumn]{article}

\usepackage[switch]{lineno} 
\usepackage{multicol,lipsum}

\usepackage[]{cite}

\usepackage{subfiles}

\usepackage{abstract}

\usepackage[table, dvipsnames]{xcolor}
\definecolor{lightgray}{gray}{0.85}

\usepackage[top=1.8cm,left=1.7cm,right=1.7cm,bottom=1.75cm]{geometry}
\usepackage{relsize}
\usepackage{tabularx}
\usepackage{array}
\usepackage{wrapfig}
\usepackage{multirow}
\usepackage{booktabs}
\usepackage{caption}
\usepackage{graphicx}
\usepackage{xcolor}
\usepackage{comment}

\usepackage[affil-it]{authblk} 
\usepackage{etoolbox}

\usepackage{lmodern}
\usepackage{amsmath}
\usepackage{amsfonts}
\usepackage{mathtools}

\makeatletter
\patchcmd{\@maketitle}{\LARGE \@title}{\fontsize{16}{19.2}\selectfont\@title}{}{}
\makeatother

\begin{document}

\title{\textbf{{Optically isotropic longitudinal piezoelectric resonant photoelastic modulator for wide angle polarization modulation at megahertz frequencies}}}

\author[1*]{Okan Atalar}
\author[1]{Amin Arbabian}

\affil[1]{\textit{Department of Electrical Engineering, Stanford University, Stanford, California 94305, USA}}
\affil[*]{Corresponding author: okan@stanford.edu\vspace{-2em}}

\date{}
\twocolumn[
  \begin{@twocolumnfalse}
\maketitle

\thispagestyle{empty}

\begin{abstract}
Polarization modulators have a broad range of applications in optics. The acceptance angle of a free-space polarization modulator is crucial for many applications. Polarization modulators that can achieve a wide acceptance angle are constructed by attaching a piezoelectric transducer to an isotropic material, and utilize a resonant transverse interaction between light and acoustic waves. Since their demonstration in the 1960s, the design of these modulators has essentially remained the same with minor improvements in the following decades. In this work, we show that a suitable single crystal with the correct crystal orientation, functioning as both the piezoelectric transducer and the acousto-optic interaction medium, could be used for constructing a highly efficient free-space resonant polarization modulator operating at megahertz frequencies and exhibiting a wide acceptance angle.  We construct the modulator using gallium arsenide, an optically isotropic and piezoelectric crystal, and demonstrate polarization modulation at 6~MHz with an input aperture of 1~cm in diameter, acceptance angle reaching $\pm30^\circ$, and modulation efficiency exceeding 50\%. Compared to state-of-the-art resonant photoelastic modulators, the modulator reported in this work exhibits greater than 50 fold improvement in modulation frequency for the same input aperture, while simultaneously reducing the thickness by approximately a factor of 80. Increasing the modulation frequency of photoelastic modulators from the kilohertz to the megahertz regime and substantially reducing their thickness lead to significant performance improvements for various use cases. This technological advancement also creates opportunities for utilizing these devices in new applications.\vspace{2em}
\end{abstract}

\end{@twocolumnfalse}
]

\section{Introduction}
Light exhibits polarization as a fundamental property, which can be modified by using a polarization modulator. This effect could be achieved through different physical mechanisms, with electro-optic and acousto-optic being the conventional modulation approaches. 

For free-space polarization modulators, large input apertures, high modulation frequency, wide acceptance angle, low-power consumption, and compact form factor is typically desired. Such modulators could find use in many settings, and single frequency modulators are used in applications ranging from polarimetry, ellipsometry, polarization spectroscopy, linear and circular dichroism, to intensity modulation of free-space beams~\cite{piezo_optical_birefringence_mod,mueller_matrix_polarimetry,pem_for_polarimetry_ellipsometry,improved_ellipsometry_pol_mod,applications_pem,stokes_and_mueller_polarimetry}. For modulation at a single frequency, resonant designs are preferred, where the drive power required to operate the polarization modulator could be substantially reduced.

Electro-optic polarization modulators constructed using materials having $\chi^{(2)}$ non-linearity (i.e., Pockels effect) offer high modulation speeds, but cannot realize large acceptance angle and low-power consumption simultaneously~\cite{enhanced_eo_ln,electro_optic_metasurface_free_space,plasmonic_phase_mod_free_space,gigahertz_mie_resonance_eo,pockels_cell_primer}. Crystals with large electro-optic coefficients typically have large birefringence (e.g., BTO, DKDP, and lithium niobate), resulting in poor acceptance angles for thick crystals since the static phase shift between the two orthogonal polarization states upon propagation through the birefringent crystal is angle dependent. Thin crystals, on the other hand, require excessive powers to operate. Optically isotropic materials that exhibit the Pockels effect (crystals belonging to the cubic system with point group $\bar{4}3$m and $23$) have significantly smaller electro-optic coupling coefficients, leading to extremely high operating powers or thick, and therefore bulky crystals for operation~\cite{eo_cubic_crystals}. Consequently, current designs are either extremely small in aperture or thick and bulky, leading to a small acceptance angle (e.g., a Pockels cell constructed using DKDP). The Kerr effect (i.e., the quadratic electro-optic effect) exists for all materials~\cite{eo_mod_handbook_optics}, however, due to the weak nature of this effect such modulators face the same problem as crystals relying on the Pockels effect; either thick crystals leading to bulky modulators or an excessive drive power for operation.

Liquid crystal electro-optic modulators offer low-power control over the polarization state of light in a compact form factor, and is the state-of-the-art technology in displays. However, they are severely limited in their modulation bandwidth to several kilohertz~\cite{LC_ultrafast,LC_1,LC_2,LC_3}. Additionally, the birefringence of liquid crystals limit the acceptance angle for such modulators. 

Acousto-optic polarization modulators are historically referred to as photoelastic modulators, and consist of an isotropic material (typically silica) carefully bonded to a suitable piezoelectric transducer. Acoustic resonance is utilized to reduce the operating power substantially. Since its conception and demonstration in the 1960s~\cite{1966_pem,improved_ellipsometry_pol_mod,piezo_optical_birefringence_mod}, the photoleastic modulator has essentially remained the same with minor improvements  and is the leading technology for achieving pure polarization modulation at a single frequency~\cite{new_design_pem,badoz1990wave,hunt1973new,diner2007dual,pem_for_polarimetry_ellipsometry,yang1995photoelastic}. Various materials functioning as the isotropic acousto-optic interaction medium have been investigated to cover different optical wavelength bands~\cite{pem_infrared,pem_infrared_2,pem_mid_infrared}. These modulators offer the advantage of a very large acceptance angle due to their lack of optical birefringence. A transverse interaction mechanism between light and acoustic waves is utilized for these modulators, resulting in a fundamental trade-off between the input aperture and modulation frequency, with current systems operating around 50~kHz with centimeter square scale apertures. Current photoelastic modulators consist of several centimeter thick isotropic bars bonded to several centimeter long piezoelectric transducers, leading to a bulky modulator. This hinders their use in applications where a small form factor is required.

Recently, photoelastic modulators using a single crystal that functions as both the piezoelectric transducer and photoelastic interaction medium have been demonstrated using lithium niobate~\cite{longitudinal_nat_paper,yz_ln_paper}. These modulators are referred to as longitudinal piezoelectric resonant photoelastic modulators (LPRPM), and utilize a collinear interaction mechanism between light and acoustic waves. This breaks the fundamental trade-off between the input aperture of the modulator and the modulation frequency, allowing simultaneously large input apertures and modulation frequencies. However, the optical birefringence of lithium niobate makes it challenging to use for applications that require pure polarization modulation. The birefringence also makes it difficult to achieve a wide acceptance angle.




In this work, we demonstrate photoelastic modulators with simultaneously large input apertures, acceptance angle, and operation in the megahertz frequency regime by constructing a LPRPM using an optically isotropic and piezoelectric material. By carefully choosing the crystal orientation of a material belonging to the cubic crystal system with point group $\bar{4}3$m or $23$, which are materials that are optically isotropic and piezoelectric, we show that it is possible to excite and confine standing shear waves in such materials. The excited standing shear waves in these modulators also couple strongly to light via the photoelastic effect for polarization modulation. 



We construct an optically isotropic LPRPM using gallium arsenide, a well developed semiconductor material exhibiting a high acousto-optic figure of merit owing to its high refractive index and large photoelastic coefficients. The modulator is fabricated using an undoped gallium arsenide wafer of thickness 234~$\mu $m and 2 inch diameter with custom crystal orientation ((332) in Miller indices notation) to optimize the relevant piezoelectric and photoelastic couplings. Indium tin oxide (ITO) is deposited on the top and bottom surfaces of the wafer to function as transparent surface electrodes. With the fabricated modulator, we demonstrate polarization modulation at approximately 6~MHz, an input aperture of 1~cm in diameter, and acceptance angle reaching $\pm30^\circ$ for 940~nm wavelength of light. Compared to state-of-the-art photoelastic modulators, this marks a greater than 50 fold improvement in modulation frequency for the same input aperture and approximately 80 times reduction in the thickness of the modulator. We demonstrate greater than 50\% modulation efficiency for a drive power of approximately 1~W for the modulator, indicating its immediate deployability for various applications. Even larger input apertures could be realized with the proposed modulator in exchange for higher operating power.


The proposed modulator offers a substantially more compact alternative to existing technology for wide acceptance angle and pure polarization modulation that finds use in many applications. Furthermore, the proposed approach opens promising avenues such as time-of-flight imaging using standard image sensors by pushing the modulation frequency from the kilohertz to the megahertz frequency regime~\cite{ToF_atalar}.

\begin{figure*}[t!]
\centering
\includegraphics[width=1\textwidth]{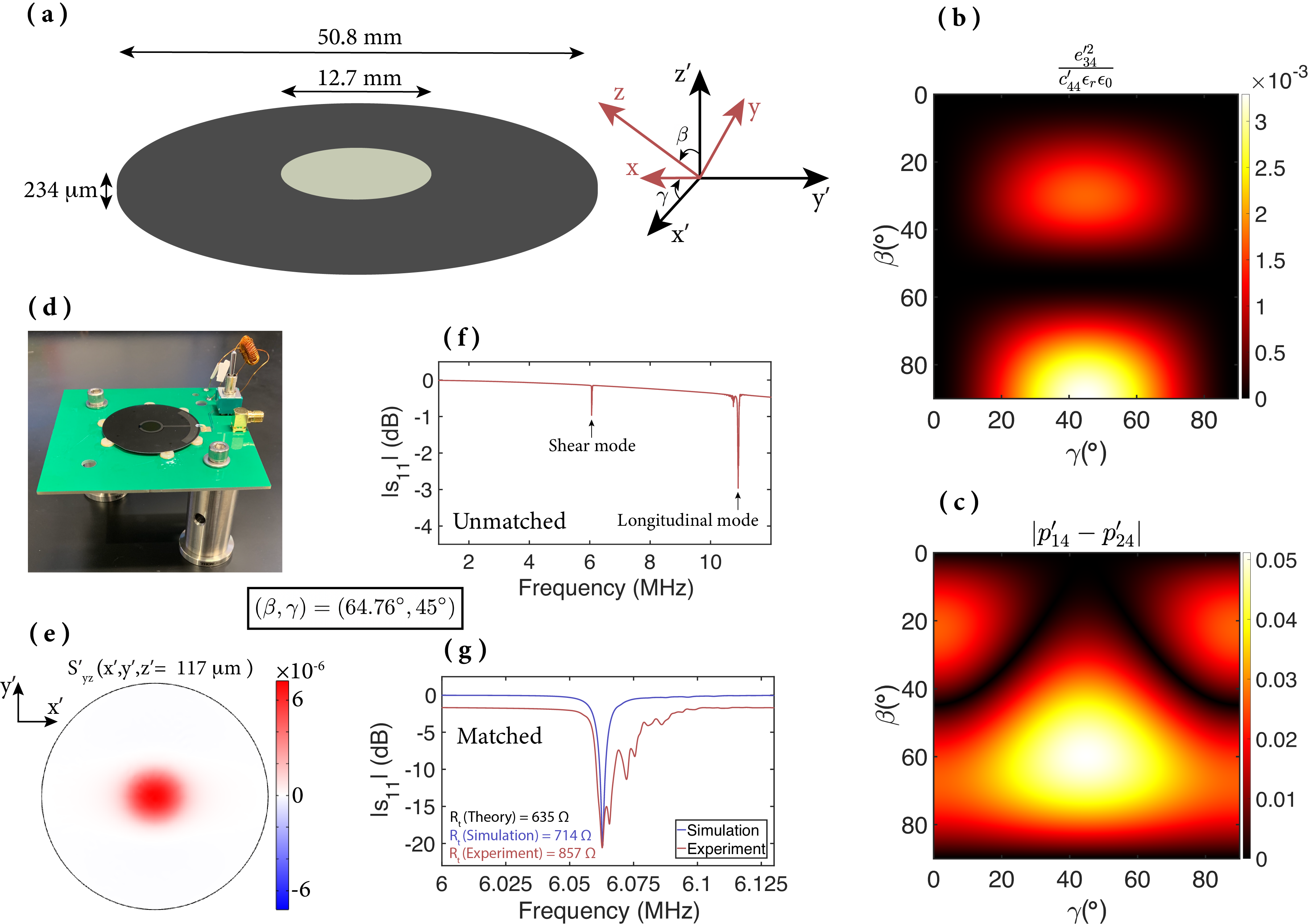}
\caption{Modulator design. (a) The modulator consists of a gallium arsenide wafer of thickness 234~$\mu $m and with a diameter of 50.8~mm coated on top and bottom surfaces with transparent surface electrodes. The surface electrodes are centered on the wafer, and have a diameter of 12.7~mm. The crystal orientation is shown with $(x,y,z)$ and the rotated coordinate system in $(x',y',z')$, where $z'$ direction is parallel to the wafer normal. The crystal coordinate system is obtained from the rotated system by applying rotations as shown, which correspond to ($\alpha = 0^{\circ}$, $\beta$, $\gamma$) rotations in Euler angles notation. (b) The electromechanical coupling coefficient corresponding to the $S'_{yz}$ shear strain is plotted as a function of the crystal orientation. (c) The absolute value of the effective photoelastic coefficient coupling $S'_{yz}$ shear strain to polarization modulation is plotted as a function of the crystal orientation. (d) Fabricated modulator wirebonded and mounted on a PCB.
(e) The simulated distribution of the dominant strain component amplitude $S'_{yz}(x',y',z' = 117~\mu \text{m})$ in the rotated coordinate frame when the wafer is excited at 6.0658~MHz with 2Vpp applied to the surface electrodes is shown for the center of the wafer. $(\beta = 64.76^{\circ}$, $\gamma = 45^{\circ})$ with a quality factor of $1.6 \times 10^3$ for this simulation. (f) Measured $|s_{11}|$ scattering parameter of the modulator is shown, where a shear resonance ($S'_{yz}$) and a longitudinal resonance ($S'_{zz}$) appear. (g) Simulated (blue) and experimental (red) of the device scattering parameter $|s_{11}|$ is shown around the shear resonance ($S'_{yz}$) when an impedance matching transformer is connected to the PCB.}
\label{fig:1}
\end{figure*}

\section{Modulator Design}
The most important choice in designing a LPRPM is the material. This influences the optical transparency window, acousto-optic figure of merit, acoustic attenuation, and piezoelectric and photoelastic tensors. Only crystals belonging to the cubic crystal system are optically isotropic (i.e., nonbirefringent), and two point groups, $\bar{4}3$m and 23, in the cubic crystal system are piezoelectric~\cite{IEEE_standard_piezoelectricity}. In these two groups, point group $\bar{4}3$m hosts promising materials: gallium phosphide and gallium arsenide. These materials exhibit some of the largest refractive indices at visible and near-infrared wavelengths~\cite{integrated_GaP}. 

In this work, we choose to use GaAs as the modulator material, as it has a large refractive index, transparency in near-infrared wavelengths, large photoelastic coefficients, and availability of custom cut wafers. With the material set to GaAs, the next step is to determine the input aperture and the modulation frequency of the modulator. To operate at megahertz frequencies, we choose to use a modulator thickness of approximately $L \approx 250 ~\mu \text{m}$, since the speed of longitudinal and shear waves in GaAs are several thousand meters per second~\cite{auld_vol1}. We fix the input aperture of the modulator to $r = 0.5~\text{cm}$ in radius. To limit clamping losses while keeping the size of the modulator small, we choose to use a wafer with a diameter of 2 inches.

The remaining step is to choose the crystal cut (i.e., crystal orientation). The cut angle of the modulator is extremely critical as it determines both the electromechanical and the acousto-optic properties of the modulator, since the single crystal functions as both the piezoelectric transducer and the photoelastic interaction medium. The chosen cut angle should accommodate coupling to the appropriate strain via piezoelectricity, while also maximizing the difference in phase modulation between the two excited eigenpolarizations in the modulator to achieve the largest polarization modulation. We choose to operate the modulator by exciting $S'_{yz}$ shear strain, as shear modes typically have lower acoustic loss compared to longitudinal modes. If we assume only $S'_{yz}$ strain is excited in the modulator, the BVD equivalent resistance is expressed in Eq.~\eqref{Eq.1} (derivation can be found in~\cite{yz_ln_paper}), where primed notation indicates representation in the rotated coordinate frame $(x',y',z')$, $(x,y,z)$ is the coordinate frame of the crystal, $c'_{44}$ is the rotated stiffness coefficient, $e'_{34}$ is the rotated piezoelectric stress constant, $f_r$ is the resonant frequency of the dominant $S'_{yz}$ strain mode, and $Q$ is the quality factor for the operating strain mode. In the primed tensor notation, $w'_{ij}$ denotes the element in the i\textsuperscript{th} row and j\textsuperscript{th} column of the rotated $\mathbf{w'}$ tensor (see Supplementary material section 2 for more details). The orientation of the crystal coordinate system with respect to the chosen rotated system is shown in Fig.~\ref{fig:1}(a). 

\begin{gather}
R_t \approx \frac{c'_{44}L}{2f_r \pi^2 r^2 e'^2_{34}Q}.  \label{Eq.1}
\end{gather}

The crystal orientation of GaAs should be selected to maximize $|p'_{14} - p'_{24}|$ (where $p'_{14}$ and $p'_{24}$ are the rotated photoelastic coefficients), while keeping the other relevant photoelastic coefficients pertaining to $S'_{yz}$ strain small, and also keeping $R_t$ at a reasonable value. The variation of the electromechanical coupling coefficient and the effective photoelastic coefficient for polarization modulation as a function of the crystal orientation in Euler angles are shown in Fig.~\ref{fig:1}(b) and Fig.~\ref{fig:1}(c), where $\epsilon_r$ is the relative permittivity of GaAs and $\epsilon_0$ is the vacuum permittivity.

The electromechanical coupling coefficient is a widely used figure of merit to asses piezoelecric transducers. This is an extremely critical parameter when constructing wide bandwidth transducers, as the transduction bandwidth is directly related to the magnitude of this coefficient. For resonant designs, however, a single frequency of operation is needed. The coupling coefficient therefore primarily influences the impedance matching condition for operation at the resonant frequency. The BVD equivalent impedance is usually selected larger than the radio frequency (RF) transmission line impedance in the design stage to make the acoustic and dielectric loss mechanisms of the modulator to dominate parasitic electrical resistances. To couple as much of the RF power carried to the modulator by a transmission line, the BVD equivalent impedance of the modulator is matched to the transmission line impedance. The electromechanical coupling coefficient should also not be too small. Due to non-ideal matching components, an extremely small electromechanical coupling coefficient would result in significant power loss. Notice that standard cuts, such as X-cut ($(\beta,\gamma) = (90^\circ,90^\circ)$), Y-cut ($(\beta,\gamma) = (90^\circ,0^\circ)$), and Z-cut ($(\beta,\gamma) = (0^\circ,0^\circ)$) do not have sufficient electromechanical coupling to be useful (see Fig.~\ref{fig:1}(b)). 

Before choosing the crystal orientation, we need to have a guess on the $Q$ for the desired mode. Based on previous devices fabricated using lithium niobate, which had $Q$ values in the range of 1,000 to 30,000, we choose the crystal orientation as $(\beta,\gamma) = (64.76^\circ,45^\circ)$, which is equivalent to (332) in Miller indices notation. The Euler angles $(\beta,\gamma) = (64.76^\circ,45^\circ)$ relate the rotated system $(x',y',z')$ to the crystal coordinate system $(x,y,z)$. This cut angle has large photoelastic coupling (as seen in Fig.~\ref{fig:1}(c)), as well as keeping $R_t$ between approximately 30~$\Omega$ and 1~k$\Omega$ for expected $Q$ values. 

With the cut angle of the wafer and its other properties determined, we perform an electromechanical simulation of the modulator using finite element modeling software COMSOL Multiphysics~\cite{COMSOL5}. The simulated dominant strain component amplitude $S'_{yz}$ for the modulator is shown in Fig.~\ref{fig:1}(e). Here, we see that a fairly uniform $S'_{yz}$ strain field can be excited in the wafer with a resonance frequency of approximately 6~MHz. The other excited strain components are negligible in comparison to the dominant $S'_{yz}$ strain (see Supplementary material section 1).



We now provide an overview of the modulation principle for the modulator. The LPRPM modulates the polarization of light via the photoelastic effect. This requires the generation of an acoustic standing wave in the wafer. Strain distribution $S'_{yz}(x',y',z')$ is generated in the active volume $V = \pi r^2 L$ of the modulator via piezoelectricity, requiring approximately $P_{RF}$ drive power, expressed in Eq.~\eqref{Eq.2} (derivation can be found in~\cite{yz_ln_paper}). 

\begin{gather}
P_{RF} \approx \frac{4 \pi f_{r} c'_{44} \int_V S'^2_{yz}(x',y',z')dV}{Q}. \label{Eq.2}
\end{gather}

The other generated strain components are negligible in comparison to $S'_{yz}(x',y',z')$ as well as having small photoelastic couplings pertinent to polarization modulation. The amplitude of polarization modulation for oblique incidence of light is expressed in spherical coordinates $(\theta,\psi)$ as Eq.~\eqref{Eq.3}, where $n$ is the refractive index of GaAs, $\lambda$ is the free-space wavelength of light, $\Delta n_{rms} (\theta,\psi,z')$ is the time-varying birefringence amplitude for the two eigenmodes of light (for the refracted light in the wafer). 

\begin{gather}
{\phi_{D}}_{rms}(\theta ,\psi) \approx \frac{2 \pi \text{cos}\big(\text{sin}^{-1}(\frac{\text{sin}\theta}{n})\big)}{\lambda} \int_{0}^L \Delta n_{rms}(\theta,\psi,z') dz'. \label{Eq.3}
\end{gather}

${\phi_{D}}_{rms}(\theta ,\psi)$ is the root mean square (rms) value of the polarization modulation amplitude evaluated over the active region of the modulator. See Supplementary material section 4 on how to calculate $\Delta n_{rms} (\theta,\psi,z')$. This analysis assumes a plane wave is incident on the modulator with incident angle $(\theta,\psi)$ with respect to the wafer normal represented in spherical coordinates.


\section{Modulator Fabrication}
We fabricate the modulator using an undoped, double-side polished GaAs wafer of thickness 234~$\mu $m, 2 inch diameter, and with crystal orientation (332) in Miller indices notation. The modulator is schematically depicted in Fig. 1(a). The wafer flat is oriented along the projection of $(\hat{a}_x - \hat{a}_y)$ to the wafer surface, expressed in the crystal coordinate frame (the wafer flat is used as a reference when orienting the transmission axis of the polarizers), where $\hat{a}_x$ and $\hat{a}_y$ are unit vectors corresponding to directions $x$ and $y$ of the crystal, respectively.

Approximately 150~nm of ITO is deposited on the top and bottom surfaces of the wafer, centered on the wafer, with a diameter of 12.7~mm using shadow masks. The ITO is deposited in a load locked chamber using sputtering. The sputtered ITO films are vacuum annealed in the chamber after deposition to improve the conductivity and transparency. 1~mm wide aluminum ring region on the edge of the ITO pattern and a microstrip region that extends to the edge of the wafer is evaporated on the top and bottom surfaces of the wafer in a load locked chamber with a thickness of 200~nm. This allows the RF signal to be carried from a printed circuit board (PCB) to the center transparent ITO region. Due to the opaqueness of aluminum deposited on the edge of the ITO pattern, the input aperture that is transparent to light is approximately 5~mm in radius. The modulator is mounted on a PCB, and the ends of the top and bottom aluminum microstrip regions are wirebonded to the PCB. The fabricated modulator mounted on a PCB is shown in Fig.~\ref{fig:1}(d).

\section{Modulator Characterization}
\subsection{Electromechanical Characterization}
We perform electromechanical characterization to observe the acoustic modes of the modulator, as well as to calculate the $Q$ for the desired mode. The characterization is performed by measuring the $s_{11}$ reflection scattering parameter with respect to 50~$\Omega$ using a vector network analyzer (VNA). A broad scan is performed first using the VNA, measuring the $s_{11}$ of the modulator from 1~MHz to 12~MHz. For this scan, 0~dBm of RF power is sent to the modulator, with a measurement bandwidth of 200~Hz and frequency stepping of 500~Hz. The magnitude of the scattering parameter is shown in Fig.~\ref{fig:1}(f), where the shear and longitudinal mode is clearly resolved. The BVD equivalent resistance $R_t$ is estimated as 867~$\Omega$ (by converting $s_{11}$ into conductance) for the desired shear mode. We attach an impedance matching transformer to match $R_t$ to 50~$\Omega$ for maximizing RF to acoustic conversion efficiency. 

Next, we perform a narrow bandwidth scan around the shear resonance with the impedance matching transformer connected to the modulator. The frequency of the VNA is swept from 6~MHz to 6.13~MHz. For this scan, 0~dBm of RF power is sent to the modulator, with a measurement bandwidth of 50~Hz and frequency stepping of 30~Hz. The magnitude of the scattering parameter for this measurement is shown in Fig.~\ref{fig:1}(g) along with the simulated $|s_{11}|$ of the modulator using COMSOL Multiphysics (acoustic damping is used to match the $Q$ of the acoustic mode in simulation with the experiment). We observe approximately 1.7~dB of loss due to the impedance matching transformer. We also see multiple acoustic modes for the experimental measurements as dips in the $|s_{11}|$. These correspond to higher order acoustic modes for the $S'_{yz}$ shear strain, and are likely a result of the distortion of the cavity (i.e., the wafer) due to clamping (we observe  changes in the $s_{11}$ of the modulator when the wafer is warped by applying loads to its edges). Here, we are interested in the acoustic mode appearing at the lowest frequency, as this mode will have the highest mode uniformity. We see a reasonable match between the simulation and experiment for the lowest order mode, where the quality factor is estimated as $Q = 1.6 \times 10^3$.

\begin{figure*}[t!]
\centering
\includegraphics[width=1\textwidth]{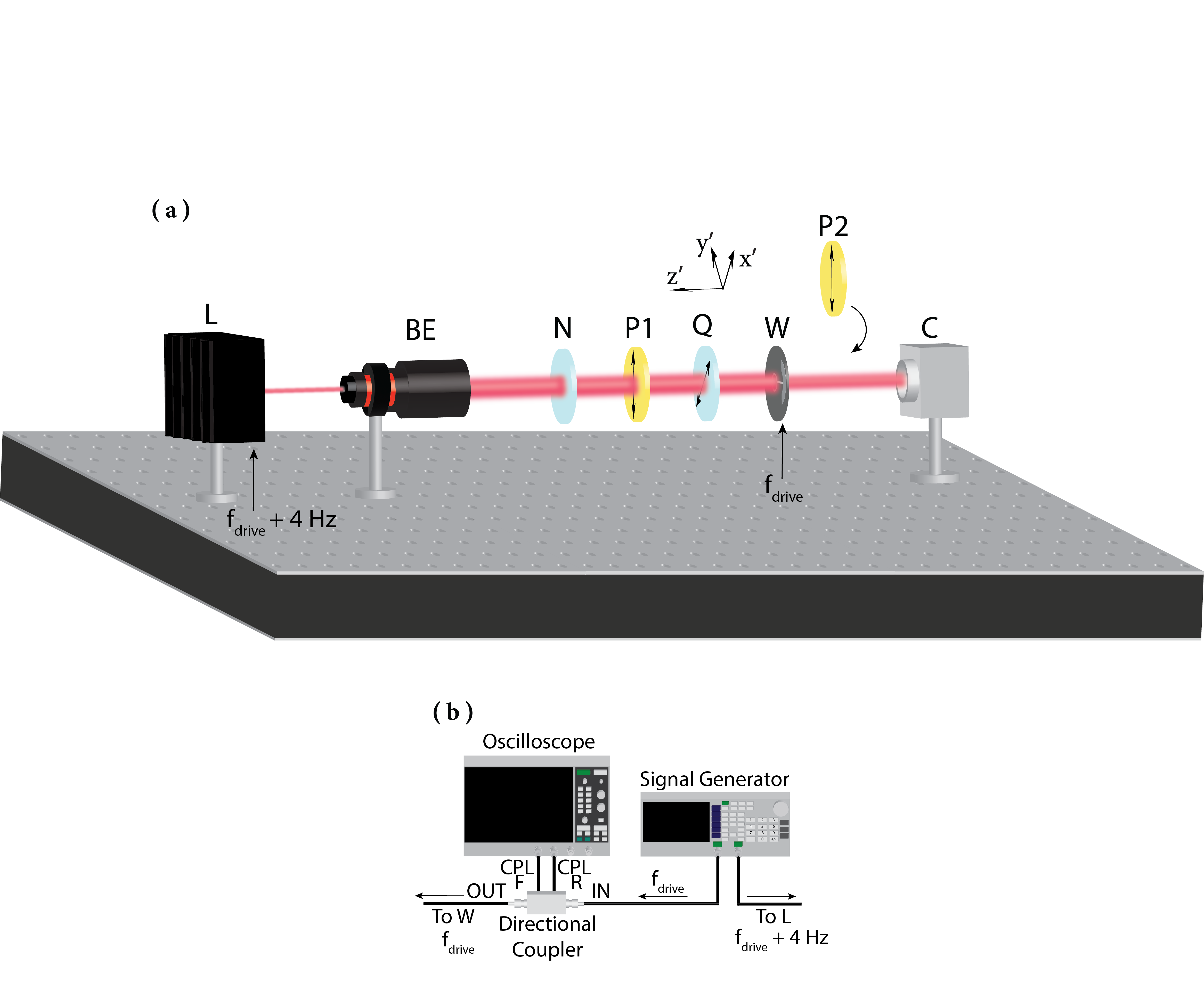}
\caption{Optical characterization setup. (a) The setup includes a laser (L) emitting light of wavelength 940~nm that is intensity modulated at $f_{drive} + 4~\text{Hz}$,
beam expander (BE) with a magnification factor of 5, neutral density filter (N), two polarizers (P1) and (P2) with transmission axis $\hat{t} = (\hat{a}'_x + \hat{a}'_y)/\sqrt{2}$, a quarter-wave plate (Q) with fast axis oriented along $\hat{a}'_x$, the modulator
(W) that is driven with an RF source of frequency $f_{drive}$, and a standard CMOS camera (C). (b) Equipment used for driving the laser (L) and the modulator (W) is shown. A directional coupler is used to send RF power to the modulator and monitor the reflected RF power with an oscilloscope.}
\label{fig:2}
\end{figure*}

\begin{figure*}[t!]
\centering
\includegraphics[width=1\textwidth]{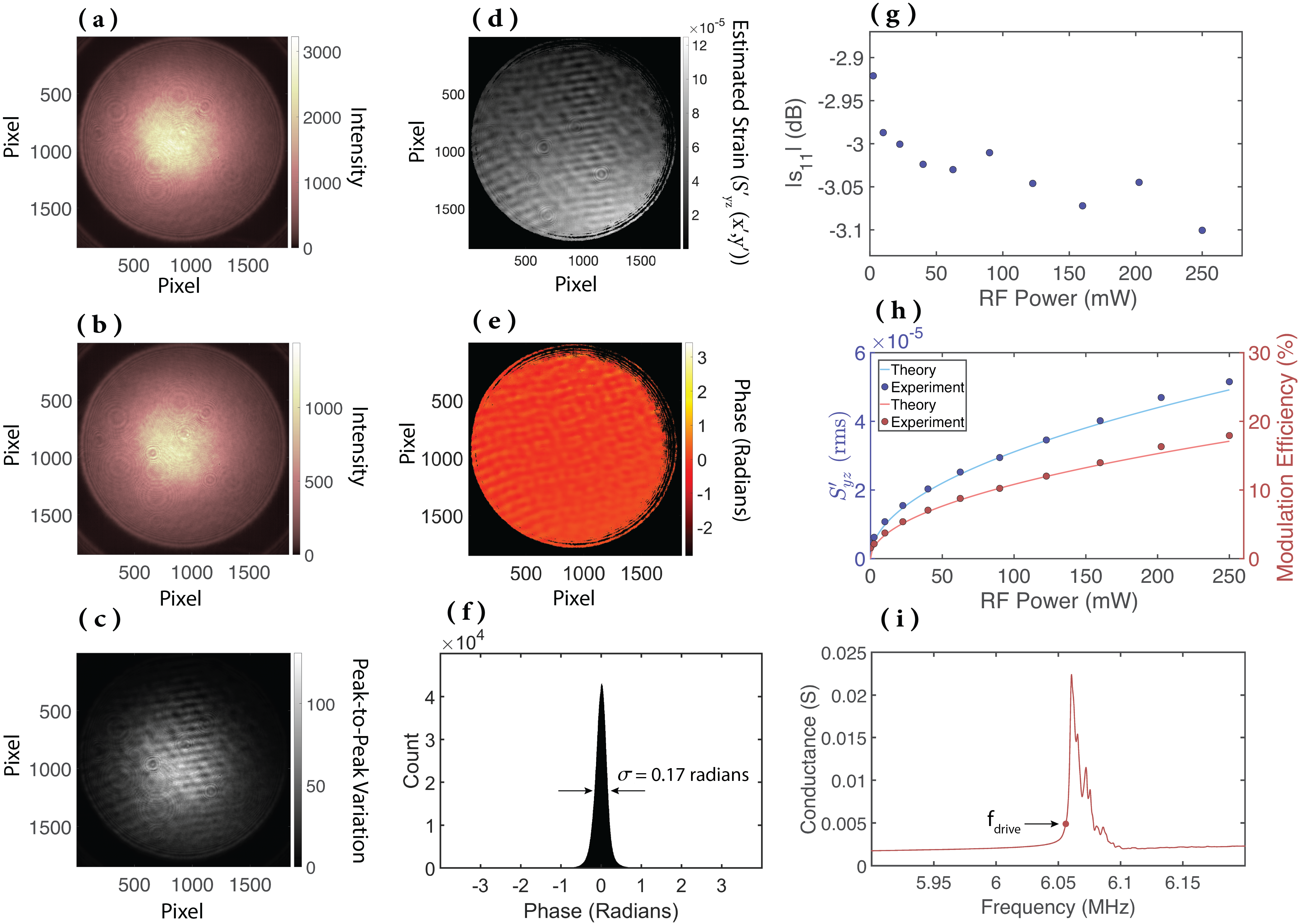}
\caption{Optical characterization results. (a) Time-averaged intensity profile of the laser beam detected by the camera per pixel is shown when the second polarizer (P2) is removed and 250~mW of RF power with frequency $f_{drive} = 6.056~\text{MHz}$ is sent to the modulator. (b) Time-averaged intensity profile of the laser beam detected by the camera per pixel is shown when the second polarizer (P2) is present and 250~mW of RF power with frequency $f_{drive} = 6.056~\text{MHz}$ is sent to the modulator. (c) The peak-to-peak variation at 4~Hz of the laser beam is shown per
pixel when the second polarizer (P2) is present and 250~mW of RF power with frequency $f_{drive} = 6.056~\text{MHz}$ is sent to the modulator. (d) Estimated strain amplitude $S'_{yz}(x',y') = \frac{\int_{0}^L S'_{yz}(x',y',z')dz'}{L}$ per pixel is shown when the second polarizer (P2) is present and 250~mW of RF power with frequency $f_{drive} = 6.056~\text{MHz}$ is sent to the modulator. (e) The phase of intensity modulation at 4~Hz of the laser beam is shown per
pixel when the second polarizer (P2) is present and 250~mW of RF power with frequency $f_{drive} = 6.056~\text{MHz}$ is sent to the modulator. (f) Histogram of the phase of intensity modulation at 4~Hz of the laser beam detected by the camera (C). The standard deviation ($\sigma$) of the phase distribution is 0.17~radians. (g) Computed $|s_{11}|$ of the modulator is shown for varying levels of RF excitation power when the drive frequency $f_{drive} = 6.056~\text{MHz}$. (h) Estimated (Experiment) and theoretical (Theory) root mean square $S'_{yz}$ and modulation efficiency evaluated over the modulator region (diameter of 1~cm) is shown for varying levels of RF excitation power. (i) Conductance of the modulator when the impedance matching transformer is used. The drive frequency of the modulator $f_{drive} = 6.056~\text{MHz}$ is shown.}
\label{fig:3}
\end{figure*}

\subsection{Optical Characterization for Normal Incidence}
We perform optical characterization to measure the profile of the dominant strain component $S'_{yz}$ excited in the modulator, as well as to measure the modulation efficiency. Perpendicular incidence of light to the wafer surface is used for this characterization. The setup shown in Fig.~\ref{fig:2} is used to optically characterize the modulator. An infrared laser diode emitting light of wavelength 940~nm and with output power of several tens of milliwatts is mounted on a laser mount. The laser beam is intensity modulated at $f_{drive} + 4~\text{Hz}$ by modulating the RF source driving the laser diode. The output beam from the laser diode is expanded by a beam expander with a magnification factor of 5 (to match the laser beam area to the active transparent electrode region of the modulator). The laser beam intensity is adjusted using a neutral density filter (to prevent it from saturating the camera pixels), and the laser beam is polarized using a polarizer with transmission axis $\hat{t} = \frac{\hat{a}'_x + \hat{a}'_y}{\sqrt{2}}$, where $\hat{a}'_x$ and $\hat{a}'_y$ are unit vectors corresponding to directions $x'$ and $y'$ in the rotated frame, respectively.

The linearly polarized light is converted into circularly polarized light after propagating through a quarter-wave plate with fast axis oriented along $\hat{a}'_x$. The light is converted into circularly polarized light to maximize the intensity modulation term at the fundamental frequency (after a polarizer is placed). The laser beam propagates through the modulator driven with RF frequency $f_{drive}$, where the light is polarization modulated at the applied RF frequency. The light finally propagates through a second polarizer with transmission axis identical to the first polarizer. The second polarizer converts the polarization modulated light into intensity modulated light. The intensity modulated light (with a beat tone at 4~Hz) is detected by a CMOS camera with megapixel spatial resolution. A second measurement is performed with the second polarizer removed from the laser beam path to calibrate the measurement. This additional measurement allows the static birefringence of the modulator (expected to be zero) and the $S'_{yz}$ strain distribution to be estimated.

The general form of the rms intensity of the laser beam after propagating through the optical elements for oblique incidence is expressed in Eq.~\eqref{Eq.4} when the second polarizer is in the laser beam path (this assumes the laser beam is not intensity modulated at the laser mount), where $c_o(\theta,\psi)$ and $c_e(\theta,\psi)$ are the amplitudes of the excited ordinary and extrordinary waves in the modulator, respectively (see Supplementary material section 4 for more details), $J_1$ is the first Bessel function of the first kind, $t$ represents time and ${I_0}_{rms}$, the rms intensity of the laser beam emitted by the laser diode (assuming the emitted laser beam is unpolarized). 

\begin{gather}
I_{rms}(\theta,\psi,t) \approx \frac{{I_0}_{rms}}{2} \Big(c_o^4(\theta,\psi) + c_e^4(\theta,\psi) \nonumber \\ - 4c_o^2(\theta,\psi)c_e^2(\theta,\psi)J_1\big({\phi_{D}}_{rms}(\theta ,\psi)\big)\text{cos}(2 \pi t f_{drive})\Big). \label{Eq.4}
\end{gather}


For this optical characterization, we intentionally drive the modulator red-detuned from the lowest order acoustic mode to minimize the mode non-uniformity. Due to the proximity of the spurious acoustic modes (see Fig.~\ref{fig:1}(g)), driving at the minimum of $|s_{11}|$ leads to coupling of substantial power into these modes. Since each mode has a different impedance, the relative phases between the excited standing waves in the modulator is non-zero, leading to non-uniformity in the phase of the excited strain distribution. This ultimately leads to non-uniform polarization modulation, where the phase imparted on the laser beam propagating through the modulator is a function of spatial position. To avoid this problem, we drive the modulator red-detuned from the lowest order acoustic mode that we would like to characterize (see Fig.~\ref{fig:3}(i)), where a compromise in the modulation efficiency is made in exchange for mode uniformity, and therefore phase uniformity. Such a driving scheme ensures that most of the RF power couples only into the desired lowest order acoustic mode, which will have the highest mode uniformity. The drive frequency is fixed to $f_{drive} = 6.056~\text{MHz}$ for this characterization to make a reasonable trade-off between the coupling of the available RF power into the modulator and the mode uniformity. 

\begin{figure*}[t!]
\centering
\includegraphics[width=1\textwidth]{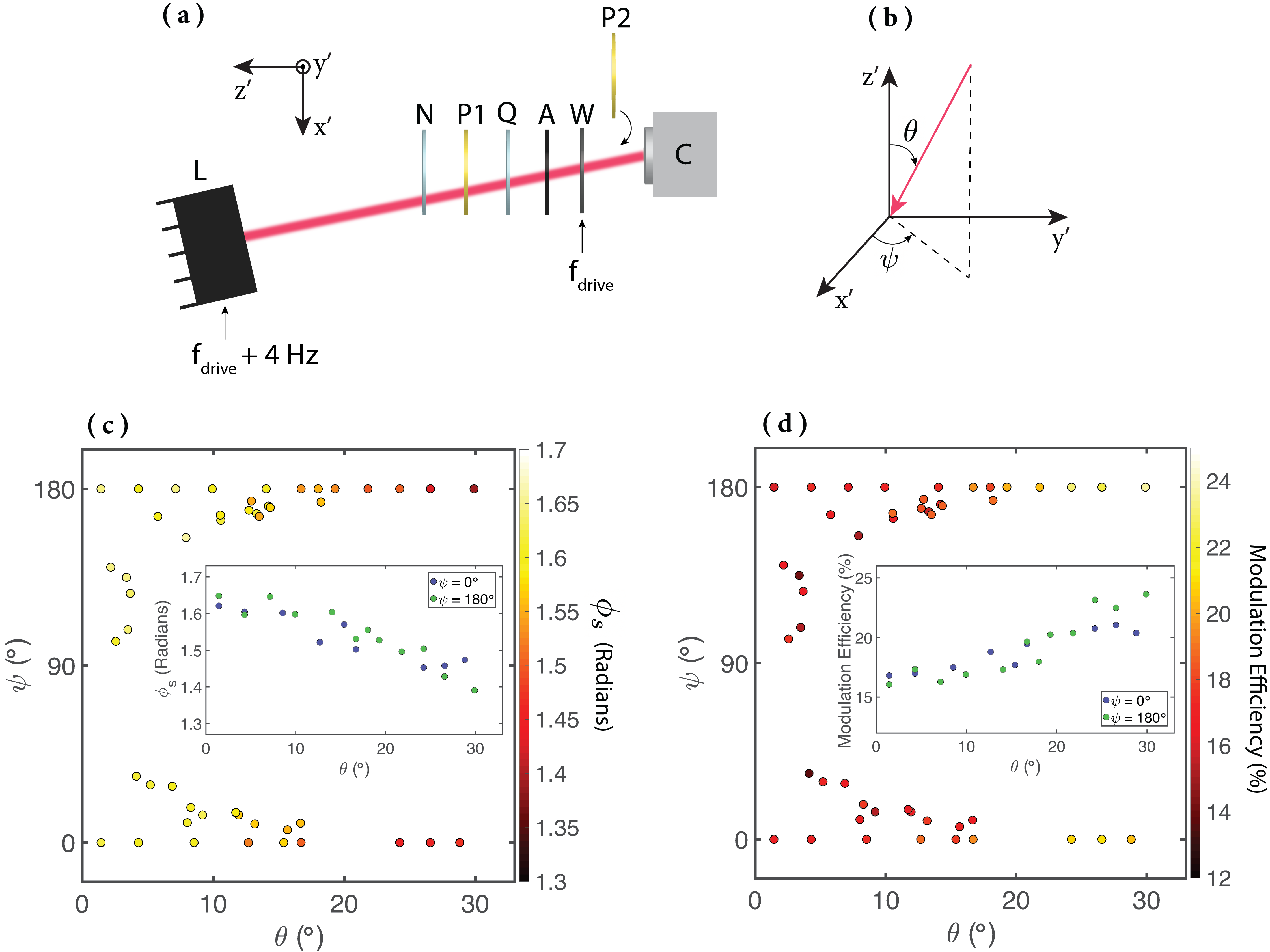}
\caption{Angular characterization setup. (a) The setup includes a laser (L) emitting light of wavelength 940~nm that is intensity modulated at $f_{drive} + 4~\text{Hz}$, neutral density filter (N), two polarizers (P1) and (P2) with transmission axis $\hat{t} = (\hat{a}'_x + \hat{a}'_y)/\sqrt{2}$, a quarter-wave plate (Q) with fast axis oriented along $\hat{a}'_x$, aperture (A) with a diameter of 2~mm, the modulator
(W) that is driven with an RF source of frequency $f_{drive} = 6.056~\text{MHz}$, and a standard CMOS camera (C). (b) The laser beam propagation direction (pink arrow) is shown in the rotated coordinate system $(x',y',z')$. (c) Estimated static birefringence $(\phi_s)$ as a function of the incoming angle of the laser beam in spherical coordinates $(\theta,\psi)$ by using the capture with and without the second polarizer (P2) in the laser beam path. (d) Estimated modulation efficiency as a function of the incoming angle of the laser beam in spherical coordinates $(\theta,\psi)$ by using the capture with and without the second polarizer (P2) in the laser beam path.}
\label{fig:4}
\end{figure*}

The optical characterization results are shown in Fig.~\ref{fig:3}, where 640 frames are captured with a frame rate of 32~Hz and exposure time of 30~$\mu $s for each frame (for both the case without and with the second polarizer in the laser beam path). The time-averaged intensity profile per pixel of the laser beam captured using the camera without and with the second polarizer are shown in Fig.~\ref{fig:3}(a) and Fig.~\ref{fig:3}(b) for 250~mW of drive power, respectively. Here, we see a reduction in the detected laser intensity by approximately a factor of 2 when the second polarizer is placed (as expected for circularly polarized light). This indicates the lack of birefringence of GaAs, as expected. The computed peak-to-peak variation per pixel at the beat tone of 4~Hz is shown in Fig.~\ref{fig:3}(c), and the extracted dominant strain amplitude per pixel $S'_{yz}(x',y') = \frac{\int_{0}^L S'_{yz}(x',y',z')dz'}{L}$ is shown in Fig.~\ref{fig:3}(d). The phase of the beat tone per pixel is shown in Fig.~\ref{fig:3}(e). This phase corresponds to the standing acoustic wave phase, and is expected (see Fig.~\ref{fig:1}(e)). The histogram of the phase distribution is shown in Fig.~\ref{fig:3}(f), with a standard deviation of 0.17~radians. Random noise sources, such as shot-noise, dark noise, and intensity noise of the laser contribute to this value. We measure the optical insertion loss of the modulator as 2.4~dB for 940~nm wavelength light. The insertion loss is due to reflection losses and absorption in the ITO film.

We also measure the variation of the rms modulation efficiency (see Supplementary material section 4), defined as the ratio in the power of the beat tone at 4~Hz to the DC level (where appropriate normalizations are taken into account) and the rms strain amplitude $S'_{yz}$ as a function of the RF drive power. The peak-to-peak variation and the estimated strain $S'_{yz}$ are averaged in rms sense over the pixels to estimate these rms values. These are plotted in Fig.~\ref{fig:3}(h) along with the theoretical values (using COMSOL Multiphysics with off-resonance drive and the relevant photoelastic coupling calculations). The $|s_{11}|$ for the modulator is tracked using the setup shown in Fig.~\ref{fig:2}(b). The $|s_{11}|$ remains between -2.9~dB and -3.1~dB for drive powers between 0~mW and 250~mW, indicating that the coupled RF power to the modulator does not vary strongly as a function of the available RF power. Since the variation of the coupled RF power to the modulator is negligible, the modulation efficiency and the rms $S'_{yz}$ should have a square root dependence on the available RF power, which we observe for the experiment as seen in Fig.~\ref{fig:3}(h). These results demonstrate that the modulator functions as expected, and the theoretical model predicts the performance of the modulator well when the operating power is in the range of 0 to 250~mW.

Hinds Instruments~\cite{Hinds_Instruments}, the leading supplier of photoelastic modulators, offers various types of modulators that cover different optical wavelength ranges. Here, we compare the performance of these modulators with our approach. We first compare in terms of the modulation frequency and the input aperture. The modulator with the highest product of modulation frequency and input aperture for visible and near-infrared light is model II/FS42, with a modulation frequency of 42~kHz and useful aperture of 27~mm. Normalizing the input aperture of model II/FS42 to 10~mm, the modulation frequency is approximately a factor of 53 greater for the GaAs modulator reported in this work.

The modulator that has the smallest thickness is model II/FS84, with a thickness of 19.05~mm. We note that this is the thickness of the packaged optical head, including the packaging of the modulator, and therefore the crystal thickness is slightly smaller. The thickness of the GaAs modulator reported in this work is approximately 81 times thinner than model II/FS84. The thickness of these commercial photoelastic modulators could be reduced in exchange for higher operating power.

\subsection{Angular Characterization}
We perform optical characterization on the modulator for different angles of incidence of the laser beam. This characterization is performed to demonstrate the wide acceptance angle of the modulator (due to its lack of optical birefringence). The characterization is similar to the optical characterization outlined in the previous section for normal incidence of light except for the positions of the optical elements and the capture of 320~frames rather than 640~frames (to speed up data capture). For this characterization, the laser mount position is changed to vary the angle of incidence of the laser beam as depicted in Fig.~\ref{fig:4}(a). The position of the laser with respect to the modulator determines the incoming angle of light, where the difference in spatial positions between the laser and the modulator are used to convert from spatial to spherical coordinates, as shown in Fig.~\ref{fig:4}(b). For each measurement at a different angle of incidence $(\theta,\psi)$, two parameters are calculated: the static birefringence ($\phi_s$) and modulation efficiency. Due to the presence of the quarter-wave plate, the static phase shift $\phi_s$ should equal $\pi/2$ radians and the modulation efficiency should not vary strongly as a function of the incidence angle of light. A liquid crystal polymer zero-order quarter-wave plate is used to ensure that the retardance imparted by the wave plate remains as close to $\pi/2$ radians for varying angles of incidence.

\begin{figure*}[t!]
\centering
\includegraphics[width=1\textwidth]{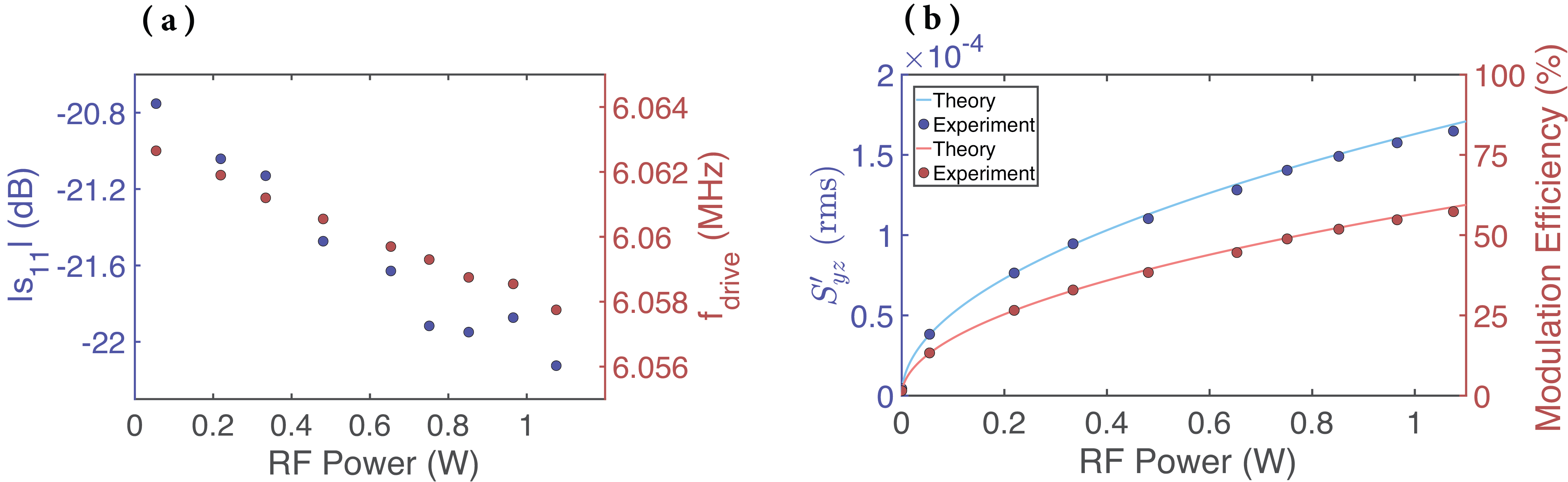}
\caption{High drive power characterization (a) Computed $|s_{11}|$ of the modulator when the drive frequency $f_{drive}$ is adjusted to minimize $|s_{11}|$ for different RF drive powers.
 (b) Estimated (Experiment) and theoretical (Theory) root mean square $S'_{yz}$ and modulation efficiency evaluated over the modulator region (diameter of 1~cm) is shown for varying levels of RF excitation power.}
\label{fig:5}
\end{figure*}

48 data points are captured, each measurement shown with a dot on the $(\theta,\psi)$ plane, where the variation of the estimated $\phi_s$ and the modulation efficiency are shown in Fig.~\ref{fig:4}(c) and Fig.~\ref{fig:4}(d), respectively.  We see that $\phi_s$ lies between 1.39~radians and 1.66~radians, and the modulation efficiency lies between 13.8\% and 23.6\% for laser incoming angle varying from approximately $-30^{\circ}$ to $30^{\circ}$. The modulation efficiency and $\phi_s$ do not vary significantly as a function of $\psi$, but vary more strongly as a function of $\theta$ (the angle between the wafer normal and the wavevector of light). 

Inset figures are plotted in Fig.~\ref{fig:4}(c) and Fig.~\ref{fig:4}(d) to demonstrate the variation of $\phi_s$ and the modulation efficiency as a function of the more relevant parameter $\theta$ (where $\psi$ is fixed). From these inset figures, we can more clearly see that the modulation efficiency increases for larger $\theta$, and $\phi_s$ decreases with increasing $\theta$. The modulation efficiency increases for larger angles of incidence $\theta$ due to the optical cavity effect of the GaAs wafer, leading to residual intensity modulation (see Supplementary material section 5). The estimated static phase shift decreases due to an increase in the transmission of the second polarizer for larger angles of incidence. Overall, the experimental results are in reasonable agreement with theory (see Supplementary material section 4), showing that we can operate the modulator with acceptance angle reaching $\pm30^\circ$.

\subsection{High Drive Power Characterization}
The attainable modulation efficiency is critical for many applications. To demonstrate high modulation efficiency, we need to generate sufficient strain in the modulator. In order to accomplish this, we drive the modulator at the minimum of its $|s_{11}|$ near the fundamental shear resonance frequency of approximately 6~MHz, and use an amplifier to increase the RF power driving the modulator. Using the setup depicted in Fig.~\ref{fig:2}(b), we drive the modulator at varying power levels and adjust the drive frequency to minimize $|s_{11}|$. The low $|s_{11}|$ coupled with small mode volume and high drive power lead to heating of the active region of the modulator. This heating leads to red-shifting of the resonance frequency. As a consequence of the high $Q$ of the modulator, the modulation efficiency drops substantially if the resonance frequency is not tracked. 

We track the drift in the resonance frequency by adjusting the drive frequency as a function of time. Every 1~second, the reflected waveform is captured with an oscilloscope using the setup depicted in Fig.~\ref{fig:2}(b). Coarse frequency stabilization is performed using the procedure outlined in~\cite{yz_ln_paper}. The coarse frequency stabilization is performed until thermal equilibrium in the modulator is reached, which takes approximately 10~seconds. After this point, the last drive frequency is fixed and used throughout the characterization. The optical characterization performed after this stage is identical to the one used for the optical characterization for normal incidence of the laser beam. The computed $|s_{11}|$ in steady-state, as well as the corresponding drive frequency is plotted in Fig.~\ref{fig:5}(a). As expected, the drive frequency $f_{drive}$ in the thermal equilibrium state gradually drops for higher RF powers (due to red-shifting of the resonance frequency). 

We observe that the $|s_{11}|$ remains below -20~dB for RF drive power up to 1.075~W, demonstrating the ability to successfully track the resonant frequency of the modulator and couple significant portion of the available RF power into the modulator. We see from Fig.~\ref{fig:5}(b) that the computed rms $S'_{yz}$ evaluated over the active region of the modulator shows the expected square root dependency on the drive power, as well as high consistency with theory taking into account elasticity and photoelasticity. This clearly shows that the modulator could be driven with sufficient power, with the linear photoelastic interaction model still holding. Modulation efficiency exceeding 50\% is demonstrated with rms strain amplitude $S'_{yz}$ exceeding $1.5 \times 10^{-4}$ in the modulator (see Supplementary material section 6 for more details). The compromise made in this driving scheme is the trade-off between modulation efficiency and the mode uniformity. The modulation efficiency increases by approximately 60\% for the same available power, but the standard deviation of the beat tone phase (which is related to the acoustic standing wave phase) increases by a factor of 4.

\section{Conclusion}
Optically isotropic and piezoelectric materials, like GaAs, can be used to construct highly efficient resonant photoelastic modulators operating at megahertz frequencies, with a large input aperture and exhibiting wide acceptance angle. In this work, we demonstrated wide angle polarization modulation at 6~MHz with an input aperture diameter of 1~cm. Acceptance angle reaching $\pm30^\circ$ and modulation efficiency exceeding 50\% for a drive power of approximately 1~W is demonstrated. Compared to state-of-the-art resonant photoelastic modulators that find use in applications ranging from polarimetry, ellipsometry, polarization spectroscopy, linear and circular dichroism, to intensity modulation of free-space beams, we report more than 50 times improvement in the modulation frequency and significant reduction in the thickness of the modulator. As future work, thinner wafers and operation using a longitudinal mode could be used to increase the modulation frequency. The impedance matching network could be improved to increase $Q$, and therefore reduce the power required to drive the modulator. The transparency window of the proposed modulator could be pushed into the visible region by using other materials in the cubic crystal system with point group $\bar{4}3$m and 23.

\section*{Funding.}
Stanford SystemX Alliance.

\section*{Acknowledgment.}
The authors thank Felix M. Mayor, Prof. Stephen E. Harris, and Prof. Amir H. Safavi-Naeini for useful discussions and Prof. Olav Solgaard for providing lab space to conduct the experiments. Part of this work was performed at the Stanford Nanofabrication Facility (SNF), supported by the National Science Foundation under award ECCS-2026822.

\section*{Disclosures.}
O.A., A.H.S.-N., and A.A. have filed a provisional patent application related to the work described in this paper.

\section*{Data availability.}
Data underlying the results presented in this paper are not publicly available at this time but may be obtained from the corresponding author upon reasonable request.

\bibliographystyle{unsrt}
\bibliography{references}

\onecolumn
\newpage


\begin{center}
  \section*{\textbf{\fontsize{16}{19.2}\selectfont Supplementary material}}
\end{center} 

\bigskip \bigskip 

\renewcommand\thefigure{S\arabic{figure}}
\setcounter{figure}{0} 

\setcounter{section}{0}


\section{Simulated complete strain profile of the modulator}
\begin{figure*}[t!]
\centering
\includegraphics[width=1\textwidth]{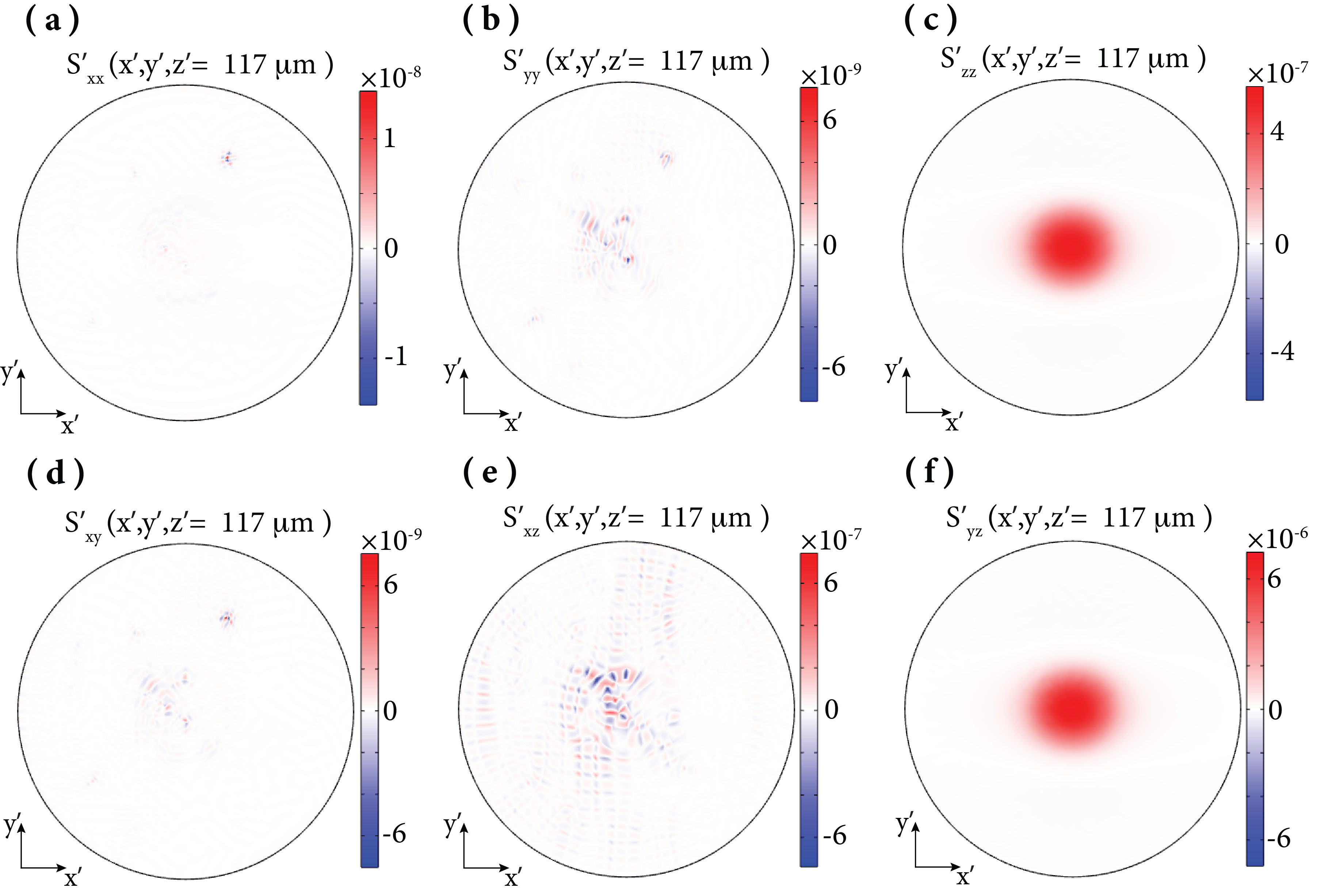}
\caption{Simulated strain amplitude profiles in the wafer. (a) $S'_{xx}$ strain amplitude profile in the rotated coordinate frame when the wafer is excited at 6.0658~MHz with 2Vpp applied to the surface electrodes is shown for the center of the wafer. (b) $S'_{yy}$ strain amplitude profile in the rotated coordinate frame when the wafer is excited at 6.0658~MHz with 2Vpp applied to the surface electrodes is shown for the center of the wafer. (c) $S'_{zz}$ strain amplitude profile in the rotated coordinate frame when the wafer is excited at 6.0658~MHz with 2Vpp applied to the surface electrodes is shown for the center of the wafer. (d) $S'_{xy}$ strain amplitude profile in the rotated coordinate frame when the wafer is excited at 6.0658~MHz with 2Vpp applied to the surface electrodes is shown for the center of the wafer. (e) $S'_{xz}$ strain amplitude profile in the rotated coordinate frame when the wafer is excited at 6.0658~MHz with 2Vpp applied to the surface electrodes is shown for the center of the wafer. (f) The dominant $S'_{yz}$ strain amplitude profile in the rotated coordinate frame when the wafer is excited at 6.0658~MHz with 2Vpp applied to the surface electrodes is shown for the center of the wafer.}
\label{fig:s1}
\end{figure*}

The complete simulated strain profiles in the GaAs wafer when excited with 2Vpp at 6.0658~MHz are shown in Fig.~\ref{fig:s1}. For this COMSOL simulation, the quality factor is $Q = 1.6 \times 10^3$, the radius of the top and bottom surface electrodes is 6.35~mm (with active area radius of $r = 5~\text{mm}$), and the wafer thickness is $ L = 234 ~\mu \text{m}$. The $S'_{yz}$ strain is the dominant strain excited in the modulator, with an amplitude more than 10 times greater than any of the other strain components.


\section{Rotated stiffness, piezoelectric, and photoelastic tensors of GaAs}
The stiffness~\cite{auld_vol1}, piezoelectric~\cite{COMSOL5}, and photoelastic~\cite{resonant_ao_coefficients} tensors are shown below for GaAs in Voigt notation, where $\mathbf{c}$ is the stiffness tensor (evaluated at constant electric field), $\mathbf{e}$ is the piezoelectric stress tensor, and $\mathbf{p}$ is the photoelastic tensor (for light wavelength of 940~nm):

\begin{gather}
\mathbf{c} = 
\begin{pmatrix}
118.8 & 59.4 & 59.4 & 0 & 0 & 0  \\
59.4 & 118.8 & 59.4 & 0 & 0 & 0 \\
59.4 & 59.4 & 118.8 & 0 & 0 & 0 \\
0 & 0 & 0 & 53.8 & 0 & 0 \\
0 & 0 & 0 & 0 & 53.8 & 0 \\
0 & 0 & 0 & 0 & 0 & 53.8 \label{Eq.S1} \tag{S1}
\end{pmatrix} \text{(GPa)}
\end{gather}

\begin{gather}
\mathbf{e} = 
\begin{pmatrix}
0 & 0 & 0 & 0.14 & 0 & 0  \\
0 & 0 & 0 & 0 & 0.14 & 0 \\
0 & 0 & 0 & 0 & 0 & 0.14 \label{Eq.S2} \tag{S2}
\end{pmatrix} (\text{Cm}^{-2})
\end{gather}

\begin{gather}
\mathbf{p} = 
\begin{pmatrix}
-0.26 & -0.285 & -0.285 & 0 & 0 & 0  \\
-0.285 & -0.26 & -0.285 & 0 & 0 & 0 \\
-0.285 & -0.285 & -0.26 & 0 & 0 & 0 \\
0 & 0 & 0 & -0.04 & 0 & 0 \\
0 & 0 & 0 & 0 & -0.04 & 0 \\
0 & 0 & 0 & 0 & 0 & -0.04 \label{Eq.S3} \tag{S3} 
\end{pmatrix} 
\end{gather}

The rotated tensors in the primed coordinate notation are related to the crystal coordinate notation through the Bond transformation matrices~\cite{bond_matrices}. The orientation of the wafer is (332) in Miller indices notation, with Euler angles $(\alpha,\beta,\gamma) = (0^\circ,64.76^\circ,45^\circ)$. The wafer flat orientation is oriented along the projection of $(\hat{a}_x - \hat{a}_y)$ to the wafer surface, expressed in the crystal coordinate frame (the primary flat is used as a reference when orienting the transmission axis of the polarizers). The Bond transformation matrices, the operations needed to transform from the crystal coordinate frame $(x,y,z)$ to the rotated coordinate frame $(x',y',z')$, and the rotated tensors are expressed below:

\begin{gather}
\mathbf{R} = 
\begin{pmatrix}
0.7071 & -0.7071 & 0 \\
0.3015 & 0.3015 & -0.9045 \\
0.6396 & 0.6396 & 0.4264 \label{Eq.S4} \tag{S4}
\end{pmatrix} 
\end{gather}

\begin{gather}
\mathbf{M} = 
\begin{pmatrix}
0.5 & 0.5 & 0 & 0 & 0 & -1  \\
0.0909 & 0.0909 & 0.8182 & -0.5455 & -0.5455 & 0.1818 \\
0.4091  & 0.4091  & 0.1818 & 0.5455 & 0.5455 & 0.8182 \\
0.1929 & 0.1929 & -0.3857 & -0.4500 & -0.4500 & 0.3857 \\
0.4523 & -0.4523 & 0 & -0.3015 & 0.3015 & 0 \\
0.2132 & -0.2132 & 0 & 0.6396 & -0.6396 & 0 \label{Eq.S5} \tag{S5}
\end{pmatrix} 
\end{gather}

\begin{gather}
\mathbf{c'} = \mathbf{M}\mathbf{c}\mathbf{M}^T = \begin{pmatrix}
142.9 & 55.02 & 39.68 & -9.3 & 0 & 0  \\
55.02 & 133.94 & 48.64 & 13.52 & 0 & 0 \\
39.68 & 48.64 & 149.27 & -4.22 & 0 & 0 \\
-9.3 & 13.52 & -4.22 & 43.04 & 0 & 0 \\
0 & 0 & 0 & 0 & 34.08 & -9.3 \\
0 & 0 & 0 & 0 & -9.3 & 49.42
\end{pmatrix} \text{(GPa)} \label{Eq.S6} \tag{S6} 
\end{gather}

\begin{gather}
\mathbf{e'} = \mathbf{R}\mathbf{e}\mathbf{M}^T = 
\begin{pmatrix}
0 & 0 & 0 & 0 & -0.0597 & 0.1266  \\
0.1266 & -0.0691 & -0.0576 & -0.0868 & 0 & 0 \\
-0.0597 & -0.0868 & 0.1465 & -0.0576 & 0 & 0 
\end{pmatrix} (\text{Cm}^{-2}) \label{Eq.S7} \tag{S7}
\end{gather}

\begin{gather}
\mathbf{p'} = \mathbf{M}\mathbf{p}\mathbf{M}^T = 
\begin{pmatrix}
-0.3125 & -0.2755 & -0.242 & 0.0202 & 0 & 0  \\
-0.2755 & -0.2930 & -0.2616 & -0.0295 & 0 & 0 \\
-0.2420 & -0.2616 & -0.3264 & 0.0092 & 0 & 0 \\
0.0202 & -0.0295 & 0.0092 & -0.0166 & 0 & 0 \\
0 & 0 & 0 & 0 & 0.0030 & 0.0202 \\
0 & 0 & 0 & 0 & 0.0202 & -0.0305
\end{pmatrix} \label{Eq.S8} \tag{S8}
\end{gather}


\section{Coupled-mode equations for normal incidence}
We assume a plane wave incident on the modulator (after propagating through the first polarizer) with normal incidence, where the input electric field of light is expressed as follows:

\begin{gather}
\bar{E}_o(\theta = 0^\circ,\psi = 0^\circ,t) = \frac{1}{2}E_oe^{j(w_Lt - kz')}\Big(\hat{a}'_x + \hat{a}'_y\Big) + \text{c.c.}\Big\rvert_{z' = 0}
\label{Eq.S9} \tag{S9}
\end{gather}

For the equation above, c.c. denotes the complex conjugate, $w_L = \frac{2 \pi c}{\lambda}$ the angular optical frequency, $j$ is the imaginary unit with $j^2 = -1$, $k = \frac{2 \pi n}{\lambda}$ the optical wavevector magnitude in GaAs. Primed notation indicates representation in the rotated coordinate system $(x',y',z')$. The output electric field is expressed as $\bar{E}_f(\theta=0^\circ,\psi=0^\circ,t)$, and the following coupled-mode equations can be arrived at using a similar derivation used in~\cite{yz_ln_paper}:

\begin{gather}
\bar{E}_f(\theta=0^\circ,\psi = 0^\circ,t) \approx \frac{1}{2}\Bigg(\hat{a}'_x\sum_{n=-\infty}^{\infty}A_n(z')e^{j((w_L + nw_r)t - kz')} + \hat{a}'_y\sum_{n=-\infty}^{\infty}B_n(z')e^{j((w_L + nw_r)t - kz')} \Bigg) + \text{c.c.}\Big\rvert_{z' = -L}
\label{Eq.S10} \tag{S10}
\end{gather}

\begin{gather}
\frac{d}{dz'} A_n(z') = \frac{-2 j n^3 p'_{14} \sqrt{\frac{\int_A S'^2_{yz}(x',y',z')dA}{\pi r^2}}  \pi \text{sin}(K z')}{\lambda}\Big(A_{n-1}(z') + A_{n+1}(z')\Big) \text{, } \forall n \in \mathbb{Z} \nonumber \\
\frac{d}{dz'} B_n(z') = \frac{-2 j n^3 p'_{24} \sqrt{\frac{\int_A S'^2_{yz}(x',y',z')dA}{\pi r^2}}  \pi \text{sin}(K z')}{\lambda}\Big(B_{n-1}(z') + B_{n+1}(z')\Big) \text{, } \forall n \in \mathbb{Z} \label{Eq.S11} \tag{S11}
\end{gather}

For the equation above, $K$ is the acoustic wavevector magnitude and $A_n(0) = B_n(0) = E_o$.

\section{Polarization modulation for oblique incidence}
In this section, we will calculate the relevant parameters for oblique incidence of light to the modulator. The derivation is similar to the one carried out in~\cite{longitudinal_nat_paper}, and solves for the case where $K = 0$ to simplify the calculations. This is referred to as the DC case, as no spatial variation is assumed for the acoustic wave (a single rms acoustic standing wave amplitude is used for the calculations). The calculation steps will be given instead of the full expressions for the relevant parameters (since the trigonometric expressions are extremely long when expressed in full form). 

We start with the wavevector direction $\hat{k} = \hat{a}'_x \text{sin}\theta \text{cos} \psi + \hat{a}'_y \text{sin}\theta \text{sin} \psi + \hat{a}'_z \text{cos}\theta$ of a plane wave represented in spherical coordinates. The transmission axis of the polarizers is $\hat{t} = \frac{\hat{a}'_x + \hat{a}'_y}{\sqrt{2}}$. The calculation of the relevant parameters are as follows:   

\begin{enumerate}
    \item The totally blocked polarization of light by the first polarizer is calculated as $\bar{p}_1(\theta,\psi) = \hat{t} \times \hat{k}$.
    \item The transmitted polarization by the polarizer is calculated as $\bar{p}_2(\theta,\psi) = \bar{p}_1(\theta,\psi) \times \hat{k}$.
    \item The plane of incidence is useful for future calculations, and is calculated as $\bar{v}(\theta,\psi) = \hat{a}'_z \times \hat{k}$.
    \item The refracted wavevector direction $\hat{k}_r(\theta,\psi)$ is found through the following relations, where $\theta_r$ is the refraction angle:

\begin{gather}
\hat{k}_r(\theta,\psi) \cdot \bar{v}(\theta,\psi) = 0 \nonumber \\
\hat{k}_r(\theta,\psi) \cdot \hat{a}'_z = \text{cos}\theta_r \nonumber \\
\text{sin}\theta = n\text{sin}\theta_r \nonumber \\
|\hat{k}_r(\theta,\psi)| = 1
\label{Eq.S12} \tag{S12}
\end{gather}
    
    \item We now find the polarizations $\bar{p}_o(\theta,\psi)$ and $\bar{p}_e(\theta,\psi)$ corresponding to the refracted eigenpolarizations. The eigenpolarizations are found by calculating the semi-axes of the ellipse formed by the intersection of the plane orthogonal to $\hat{k}_r(\theta,\psi)$ and the modified index ellipsoid via photoelasticity. The time-varying index ellipsoid for GaAs as a function of spatial position is expressed as follows, where $f_{drive}$ is the excitation frequency of the RF source applied to the modulator through the surface electrodes:

    \begin{gather}
    x'^2\Big(\frac{1}{n^2} + 2p'_{14}S'_{yz}(x',y',z')\text{cos}(2 \pi t f_{drive})\Big) + y'^2\Big(\frac{1}{n^2} + 2p'_{24}S'_{yz}(x',y',z')\text{cos}(2 \pi t f_{drive})\Big) \nonumber \\ + z'^2\Big(\frac{1}{n^2} + 2p'_{34}S'_{yz}(x',y',z')\text{cos}(2 \pi t f_{drive})\Big) + 2y'z'\Big(\frac{1}{n^2} + 2p'_{44}S'_{yz}(x',y',z')\text{cos}(2 \pi t f_{drive})\Big) = 1 \label{Eq.S13} \tag{S13}
    \end{gather}

    The vectors pointing from the origin to the semi-axes of the ellipsoid correspond to the polarizations $\bar{p}_o(\theta,\psi)$ and $\bar{p}_e(\theta,\psi)$, respectively. The lengths of the axes correspond to the time-varying refractive indices experienced by the ordinary $n_o(\theta,\psi,x',y',z',t)$ and extraordinary wave $n_e(\theta,\psi,x',y',z',t)$, respectively.

    \item The time-varying birefringence due to photoelasticity is calculated as $\Delta n(\theta,\psi,x',y',z',t) = n_o(\theta,\psi,x',y',z',t) - n_e(\theta,\psi,x',y',z',t)$. The root mean square (rms) birefringence is expressed as follows, where the surface integral is carried out over the aperture of the modulator (with area $\pi r^2$): 

    \begin{gather}
    \Delta n_{rms} (\theta,\psi,z',t) = \sqrt{\frac{\int_A \Delta n^2(\theta,\psi,x',y',z',t)dA}{\pi r^2}} \label{Eq.S14} \tag{S14}
    \end{gather}

    The amplitude of the time-varying birefringence is expressed as follows:

\begin{gather}
\Delta n_{rms} (\theta,\psi,z',t) = \Delta n_{rms} (\theta,\psi,z')\text{cos}(2 \pi t f_{drive})
\label{Eq.S15} \tag{S15}
\end{gather}

$\Delta n_{rms} (\theta,\psi,z')$ converges to its expected value for normal incidence of light $(\hat{k} = \hat{a}'_z)$ (corresponding to polarization modulation of light).

\begin{gather}
\lim_{(\theta,\psi) \to (0^\circ,0^\circ)} \Delta n_{rms} (\theta,\psi,z') = n^3 \big(p'_{24} - p'_{14}\big)\sqrt{\frac{\int_A S'^2_{yz}(x',y',z')dA}{\pi r^2}}
\label{Eq.S16} \tag{S16}
\end{gather}

\item We now calculate the amplitudes of the excited ordinary ($c_o(\theta,\psi)$) and extraordinary ($c_e(\theta,\psi)$) polarizations in the modulator (for transmitted polarization $\bar{p}_2(\theta,\psi)$ through the first polarizer). These amplitudes are found by calculating the inner product of the refracted polarization $\bar{p}_r(\theta,\psi)$ and the refracted eigenpolarizations $\bar{p}_o(\theta,\psi)$ and $\bar{p}_e(\theta,\psi)$. To simplify the analysis, we assume that no reflection occurs at the air-modulator interface. 

\item We now calculate the polarization $\bar{p}_r(\theta,\psi)$ corresponding to the refracted light. Using the plane of incidence $\bar{v}(\theta,\psi)$, we analyze the p polarization (polarization parallel to the plane of incidence) and the s polarization (polarization perpendicular to the plane of incidence) separately. We express $\bar{p}_2(\theta,\psi)$ in the basis of the p ($\bar{p}_{2p}(\theta,\psi)$) and s ($\bar{p}_{2s}(\theta,\psi)$) polarizations. These polarizations satisfy the following relations:

\begin{gather}
\bar{p}_{2p}(\theta,\psi) \cdot \hat{k} = 0 \nonumber \\
\bar{p}_{2s}(\theta,\psi) \cdot \hat{k} = 0 \nonumber \\
\bar{p}_{2p}(\theta,\psi) \cdot \hat{v}(\theta,\psi) = 0 \nonumber \\
\bar{p}_{2p}(\theta,\psi) \cdot \bar{p}_{2s}(\theta,\psi) = 0 \label{Eq.S17} \tag{S17}
\end{gather}

The two polarizations are found as follows:

\begin{gather}
\bar{p}_{2s}(\theta,\psi) = \bar{p}_2(\theta,\psi) - \Big(\bar{p}_2(\theta,\psi) \cdot  \big(\hat{k} \times \hat{v}(\theta,\psi)\big)\Big) \Big(\hat{k} \times \hat{v}(\theta,\psi)\Big) \nonumber \\
\bar{p}_{2p}(\theta,\psi) = \bar{p}_2(\theta,\psi) - \bar{p}_{2s}(\theta,\psi) \label{Eq.S18} \tag{S18}
\end{gather}

We now relate these s and p polarizations in air to the refracted s ($\bar{p}_{rs}(\theta,\psi)$) and p ($\bar{p}_{rp}(\theta,\psi)$) polarizations in GaAs. The relations are as follows:

\begin{gather}
\bar{p}_{rs}(\theta,\psi) = \bar{p}_{2s}(\theta,\psi) \nonumber \\
\bar{p}_{rp}(\theta,\psi) = |\bar{p}_{2p}(\theta,\psi)|\Big(\hat{k}_r(\theta,\psi) \times \hat{p}_{rs}(\theta,\psi)\Big) \label{Eq.S19} \tag{S19}
\end{gather}

The refracted polarization is finally found as $\bar{p}_{r}(\theta,\psi) = \bar{p}_{rs}(\theta,\psi) + \bar{p}_{rp}(\theta,\psi)$.

\item The excited ordinary and extraordinary wave amplitudes are calculated as follows:

\begin{gather}
c_o(\theta,\psi) = \bar{p}_r(\theta,\psi) \cdot \bar{p}_o(\theta,\psi) \nonumber \\
c_e(\theta,\psi) = \bar{p}_r(\theta,\psi) \cdot \bar{p}_e(\theta,\psi) \label{Eq.S20} \tag{S20}
\end{gather}

\end{enumerate}

After propagating through the second polarizer, the intensity of the plane wave with wavevector direction $\hat{k}$ is expressed as in Eq. 5 in the main manuscript (including a quater-wave plate placed between the modulator and the first polarizer to impart a $\pi/2$ radians phase shift between the ordinary and extraordinary polarizations). The depth of modulation ($DOM(\theta,\psi)$) is defined as follows (the ratio of the beat tone amplitude to the DC value):

\begin{gather}
DOM(\theta,\psi) = \frac{4 c_o^2(\theta,\psi) c_e^2(\theta,\psi) J_1\big({\phi_{D}}_{rms}(\theta,\psi)\big)}{c_o^4(\theta,\psi) + c_e^4(\theta,\psi)} \label{Eq.S21} \tag{S21}
\end{gather}

\begin{figure*}[t!]
\centering
\includegraphics[width=0.75\textwidth]{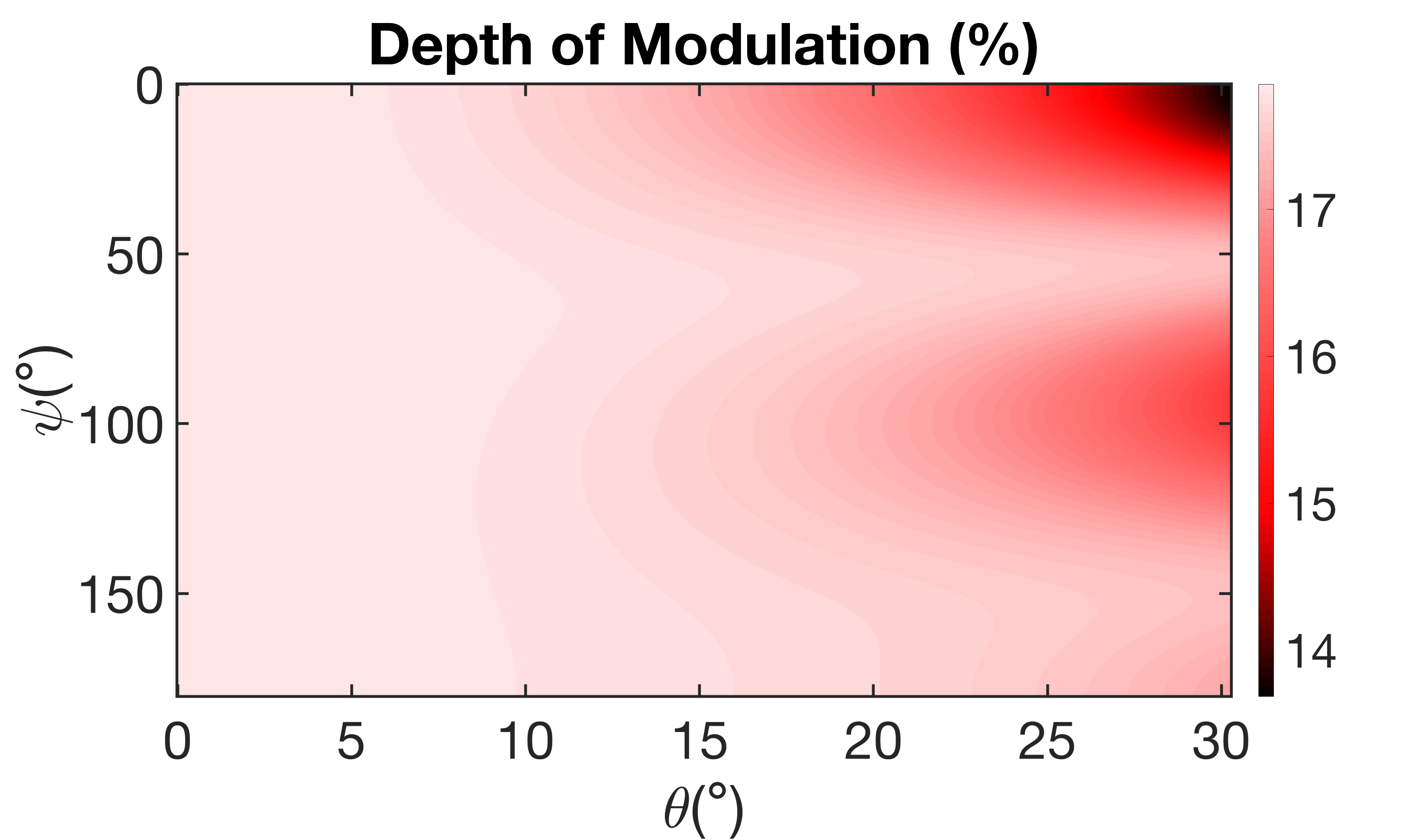}
\caption{Depth of modulation ($DOM(\theta,\psi)$) plotted as a function of the incidence angle of light for rms strain amplitude $S'_{yz} = 5.15 \times 10^{-5}$.}
\label{fig:s2}
\end{figure*}

The plot of an approximate form of $DOM(\theta,\psi)$ as a function of the incidence angle is shown in Fig.~\ref{fig:s2} for rms strain amplitude $S'_{yz} = 5.15 \times 10^{-5}$. We observe that the variance is relatively small for angle of incidence reaching up to $30^\circ$ (and therefore the acceptance angle is large), as expected (since GaAs is optically isotropic, and therefore we do not see nulls caused by birefringence). 

\section{Residual intensity modulation}
Most of the light that is incident on the modulator passes through the modulator once. Some portion of the light, however, bounces multiple times before exiting the wafer (due to the large refractive index of GaAs). The interference of these two (or more) paths that have propagated a different distance to the detector leads to intensity modulation (in addition to intensity modulation due to polarization modulation and the polarizers). 

\begin{figure*}[t!]
\centering
\includegraphics[width=0.75\textwidth]{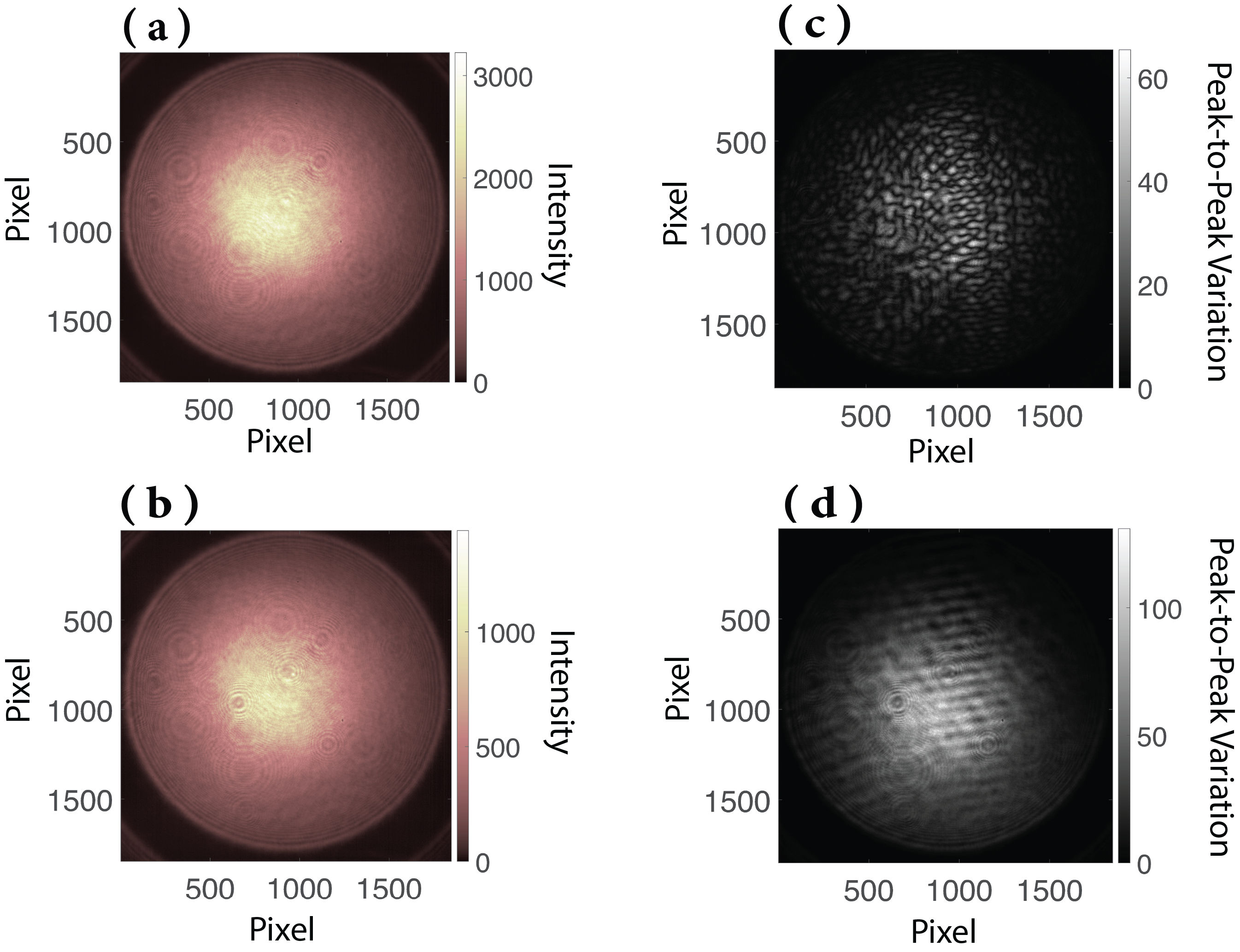}
\caption{Optical measurements with and without the second polarizer (P2). (a) Time-averaged intensity profile of the laser beam detected by the camera per pixel is shown when the second polarizer (P2) is removed and 250~mW of RF power with frequency $f_{drive} = 6.056~\text{MHz}$ is sent to the modulator. (b) Time-averaged intensity profile of the laser beam detected by the camera per pixel is shown when the second polarizer (P2) is present and 250~mW of RF power with frequency $f_{drive} = 6.056~\text{MHz}$ is sent to the modulator. (c) The peak-to-peak variation at 4~Hz of the laser beam is shown per
pixel when the second polarizer (P2) is removed and 250~mW of RF power with frequency $f_{drive} = 6.056~\text{MHz}$ is sent to the modulator. (d) The peak-to-peak variation at 4~Hz of the laser beam is shown per
pixel when the second polarizer (P2) is present and 250~mW of RF power with frequency $f_{drive} = 6.056~\text{MHz}$ is sent to the modulator.}
\label{fig:s3}
\end{figure*}

We have performed a measurement for normal incidence of light with and without the second polarizer present in the beam path for the results corresponding to Fig. 3 in the main manuscript. The optical characterization results to investigate this effect are shown in Fig.~\ref{fig:s3}. The rms peak-to-peak variation at the beat tone of 4~Hz is calculated for the case with and without the second polarizer present in the laser beam path. Accounting for the intensity difference of the laser beam with and without the second polarizer present, the rms beat tone amplitude for the case with the second polarizer present is more than a factor of 8 larger than the case without the polarizer. We can therefore conclude that the dominant contributor to intensity modulation is via polarization modulation. For oblique incidence of light, larger incidence angles typically result in larger reflection coefficients, leading to higher residual intensity modulation. This effect is observed in Fig. 4 in the main manuscript, where the computed modulation efficiency increases for increasing $\theta$ (angle of incidence with respect to the wafer normal).

\section{High RF power optical characterization when modulator is driven at the minimum of $|s_{11}|$}
\begin{figure*}[t!]
\centering
\includegraphics[width=1\textwidth]{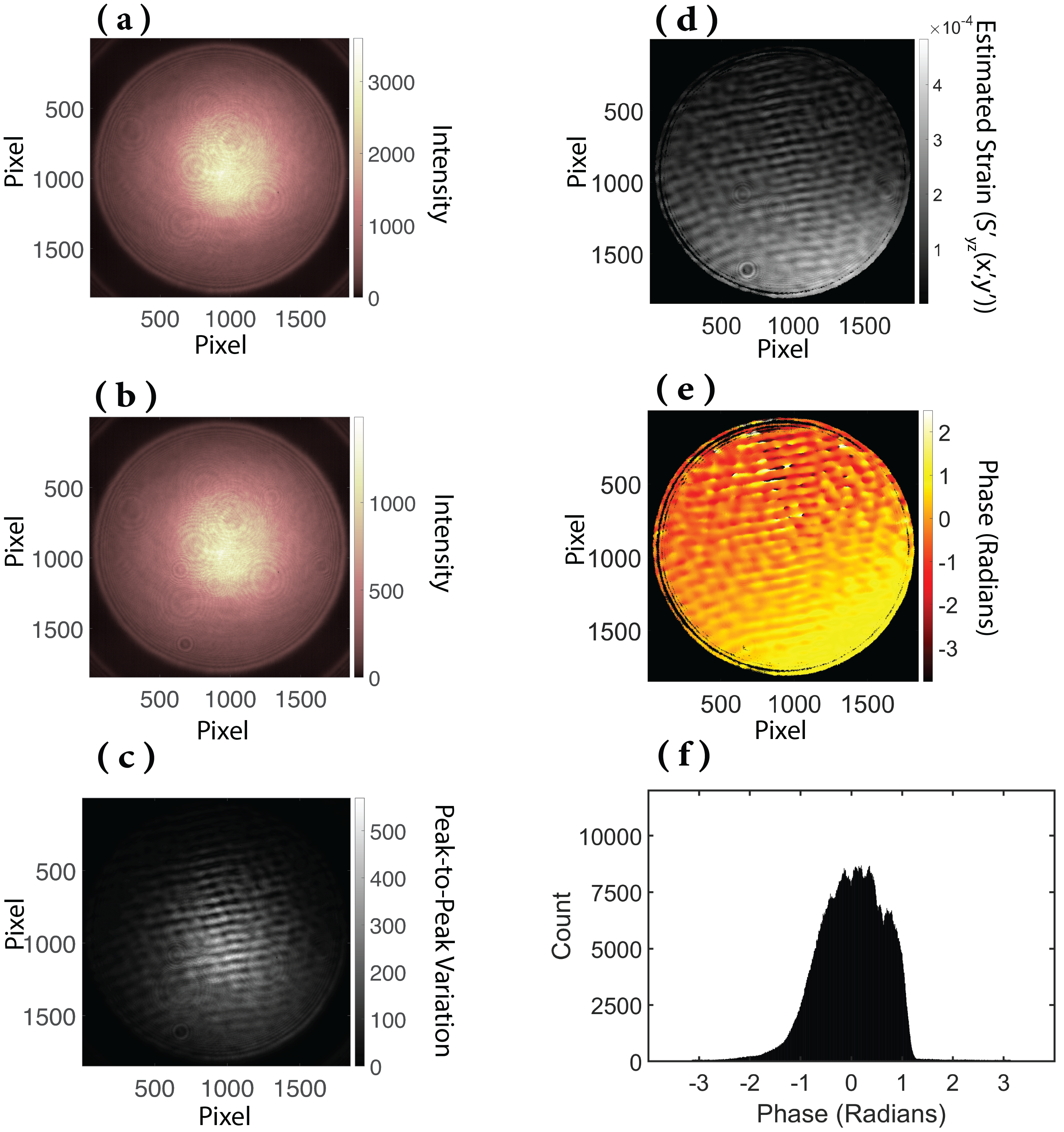}
\caption{Optical characterization results when the modulator is driven at the minimum of its $|s_{11}|$. (a) Time-averaged intensity profile of the laser beam detected by the camera per pixel is shown when the second polarizer (P2) is removed and 1.075~W of RF power with frequency $f_{drive} = 6.05775~\text{MHz}$ is sent to the modulator. (b) Time-averaged intensity profile of the laser beam detected by the camera per pixel is shown when the second polarizer (P2) is present and 1.075~W of RF power with frequency $f_{drive} = 6.05775~\text{MHz}$ is sent to the modulator. (c) The peak-to-peak variation at 4~Hz of the laser beam is shown per
pixel when the second polarizer (P2) is present and 1.075~W of RF power with frequency $f_{drive} = 6.05775~\text{MHz}$ is sent to the modulator. (d) Estimated strain amplitude $S'_{yz}(x',y') = \frac{\int_{0}^L S'_{yz}(x',y',z')dz'}{L}$ per pixel is shown when the second polarizer (P2) is present and 1.075~W of RF power with frequency $f_{drive} = 6.05775~\text{MHz}$ is sent to the modulator. (e) The phase of intensity modulation at 4~Hz of the laser beam is shown per
pixel when the second polarizer (P2) is present and 1.075~W of RF power with frequency $f_{drive} = 6.05775~\text{MHz}$ is sent to the modulator. (f) Histogram of the phase of intensity modulation at 4~Hz of the laser beam detected by the camera (C). The standard deviation ($\sigma$) of the phase distribution is 0.68~radians.}
\label{fig:s4}
\end{figure*}

Similar to the optical characterization performed when the modulator was driven at $f_{drive} = 6.056~\text{MHz}$, we perform an optical characterization when the modulator is driven at the minimum of its $|s_{11}|$ near the shear resonance frequency of approximately 6~MHz. The optical characterization results are shown in Fig.~\ref{fig:s4}. Here, we drive the modulator with high power using an amplifier, sending $1.075~\text{W}$ of RF power. Due to high drive power, the wafer heats up, leading to a red shift in the resonant frequency. Before the optical characterization is performed, a crude frequency stabilization step is used (see Supplementary Information section of~\cite{yz_ln_paper} for more details), where the drive frequency is adjusted at each time step to track the resonance. Once thermal equilibrium is reached in the modulator, the drive frequency $f_{drive}$ is maintained, and this frequency is used during the optical characterization data capture.

As seen in Fig.~\ref{fig:s4}(f), the phase distribution of the beat tone has a large variance of approximately 0.68~radians. The large variance of the detected phase is a consequence of the non-uniformity of the strain phase (which is likely due to coupling of significant power into spurious acoustic modes). 

\begin{figure*}[t!]
\centering
\includegraphics[width=1\textwidth]{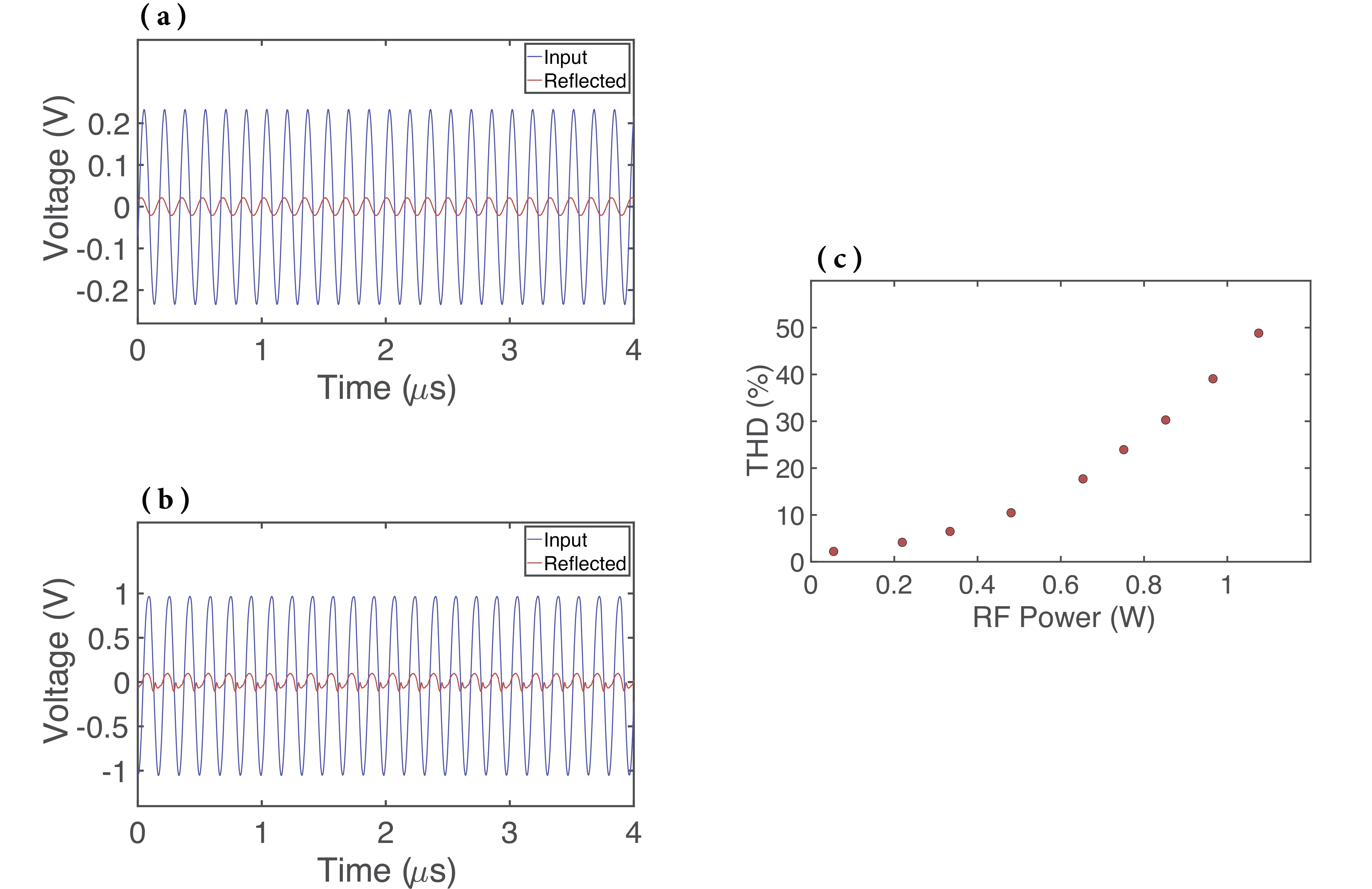}
\caption{Acoustic non-linearity at high RF drive power. (a) Input (sent) and reflected RF voltage as a function of time when 54.3~mW of RF power is sent to the modulator with frequency $f_{drive} = 6.06265~\text{MHz}$. (b) Input (sent) and reflected RF voltage as a function of time when 1.075~W of RF power is sent to the modulator with frequency $f_{drive} = 6.05775~\text{MHz}$. (c) Total harmonic distortion (THD) of the reflected waveform from the modulator as a function of the RF drive power when the modulator is driven at the minimum of its $|s_{11}|$.}
\label{fig:s5}
\end{figure*}

When the modulator is driven with sufficiently high drive power, and when the $|s_{11}|$ is small (here, it is less than -20~dB), sufficient strain builds up in the wafer that acoustic non-linearities start becoming noticeable. Using the setup depicted in Fig.~2(b) in the main manuscript, we probe the sent and reflected RF power from the modulator. The results are shown in Fig.~\ref{fig:s5}, where we see noticeable distortion of the received waveform, caused by acoustic non-linearities. As seen in Fig.~\ref{fig:s5}(b), sufficient strain is excited in the wafer that acoustic non-linearities are noticeable. However, the optical modulation still remains sufficiently linear as seen in Fig.~5(b) in the main manuscript.

\end{document}